\pdfoutput=1
\documentclass[12pt]{article}
\usepackage[a4paper,margin=25mm]{geometry}
\usepackage{graphicx}
\usepackage{amsmath,amssymb}
\usepackage{placeins,float}
\usepackage{flafter}
\usepackage{hyperref}
\newif\ifarxiv
\arxivtrue
\newif\ifseparatesupplement
\separatesupplementtrue
\usepackage{comment}

\newcommand{\PaperKeywords}{all-optical switching, GdFeCo, magneto-optic
readout, Faraday rotation, integrated photonics, magnetophotonics,
finite-difference time-domain, Landau--Lifshitz--Bloch model}

\newcommand{\OrcidBudi}{0000-0001-8393-6282}
\newcommand{\OrcidAyi}{0000-0002-0275-7846}

\ifarxiv
  \includecomment{expository}
  \excludecomment{brief}
  \includecomment{fullsection}
  \ifseparatesupplement
    \includecomment{stub}
  \else
    \excludecomment{stub}
  \fi
\else
  \excludecomment{expository}
  \includecomment{brief}
  \excludecomment{fullsection}
  \includecomment{stub}
\fi

\ifarxiv
  \newcommand{\FigW}[1]{#1\textwidth}
\else
  \newcommand{\FigW}[1]{\textwidth}
\fi

\ifarxiv
  \ifseparatesupplement

    \newcommand{\Secens}{Supplementary~S2}
    \newcommand{\Figmesh}{figure~S1.2}

    \newcommand{\SecheatingC}{Supplementary~S3}

  \else

    \newcommand{\Secens}{section~\ref{sec:ensemble}}
    \newcommand{\Figmesh}{figure~\ref{fig:mesh}}

    \newcommand{\SecheatingC}{Section~\ref{sec:heating}}

  \fi
\else

  \newcommand{\Secens}{Supplementary~S2}
  \newcommand{\Figmesh}{figure~S1.2}

  \newcommand{\SecheatingC}{Supplementary~S3}

\fi
\newcommand{\PackageVersion}{0.2.0}

\newcommand{\ForcedSwitchFraction}{0.782}
\newcommand{\ToggleSwitchFraction}{0.859}
\newcommand{\ToggleSwitchPercent}{86\%}
\newcommand{\LatchDecidedFraction}{0.876}
\newcommand{\FeCoCrossing}{\ensuremath{0.86\,\mathrm{ps}}}
\newcommand{\GdCrossing}{\ensuremath{2.15\,\mathrm{ps}}}
\newcommand{\SublatticeDelay}{\ensuremath{1.29\,\mathrm{ps}}}
\newcommand{\PerCellSaturation}{0.990}

\newcommand{\FilmAverageToggle}{\ensuremath{-0.711}}

\newcommand{\CompensationTemperature}{\ensuremath{250\,\mathrm{K}}}

\newcommand{\TePeak}{\ensuremath{2187\,\mathrm{K}}}

\newcommand{\CurieTemperature}{\ensuremath{540\,\mathrm{K}}}
\newcommand{\ProductionFluence}{\ensuremath{0.813\,\mathrm{mJ\,cm^{-2}}}}

\newcommand{\SeedCount}{16}
\newcommand{\ActiveCellCount}{1323}

\newcommand{\OccupancyMean}{\ensuremath{0.502 \pm 0.004}}

\newcommand{\PairCount}{120}

\newcommand{\NoiseSigmaTM}{\ensuremath{1.2\,\mathrm{T}}}
\newcommand{\NoiseSigmaRE}{\ensuremath{1.6\,\mathrm{T}}}
\newcommand{\NoiseNormDeviation}{\ensuremath{1.8\times10^{-7}}}

\newcommand{\FluenceThreshold}{\ensuremath{0.176\,\mathrm{mJ\,cm^{-2}}}}

\newcommand{\FluenceWindowRange}{0.62--1.49}
\newcommand{\FluenceWindowFactor}{2.4}
\newcommand{\AnchorSwitchRange}{0.81--0.87}

\newcommand{\ConvergenceOrderDielectric}{1.98}
\newcommand{\ConvergenceRSquaredDielectric}{0.997}
\newcommand{\ConvergenceOrderDispersive}{1.0}
\newcommand{\ConvergenceRSquaredDispersive}{0.74}

\newcommand{\FaradayPedestal}{\ensuremath{+0.12^{\circ}}}

\newcommand{\FaradaySatOdd}{\ensuremath{1.28^{\circ}}}
\newcommand{\FaradayMZeroOdd}{\ensuremath{0.107^{\circ}}}
\newcommand{\FaradaySuppression}{12}
\newcommand{\KerrPedestal}{\ensuremath{-11.77^{\circ}}}

\newcommand{\KerrSatOdd}{\ensuremath{8.24^{\circ}}}
\newcommand{\KerrMZeroResidual}{\ensuremath{0.063^{\circ}}}
\newcommand{\KerrSuppression}{130}

\newcommand{\XYPeakRatio}{1.99}

\newcommand{\LobeBalancePlus}{1.48}
\newcommand{\LobeBalanceZero}{0.96}
\newcommand{\TransmittedParity}{\ensuremath{4\%}}

\title{Multiphysics simulation of all-optical switching and sequential magneto-optic readout in an integrated Si$_3$N$_4$/GdFeCo waveguide}
\author{Muhammad Arief Mulyana, Budi Adiperdana$^{\ast}$ and Ayi Bahtiar\\[2pt]
\normalsize Department of Physics, Faculty of Mathematics and Natural Sciences,\\
\normalsize Universitas Padjadjaran, Jl.~Ir.~Soekarno Km.~21,\\
\normalsize Jatinangor, Sumedang 45363, West Java, Indonesia\\[2pt]
\normalsize $^{\ast}$Corresponding author: \texttt{budi.adiperdana@phys.unpad.ac.id}\\[2pt]
\normalsize ORCID:\ B.~Adiperdana \href{https://orcid.org/\OrcidBudi}{\OrcidBudi};\
\normalsize A.~Bahtiar \href{https://orcid.org/\OrcidAyi}{\OrcidAyi}}
\date{September 2026}

\begin{document}
\maketitle

\begin{abstract}
Femtosecond pulses guided in a silicon nitride waveguide have recently switched a Co/Gd magnetic
element patterned on the waveguide, a step towards magnetic memory integrated with photonic
circuits. The guided mode heats such an element non-uniformly, and, to our knowledge, no model has
yet followed the written magnetic state from this heating to an optical readout. Here we simulate
both writing and readout for an 8 nm GdFeCo film inserted across a Si$_3$N$_4$ waveguide.
Three-dimensional finite-difference time-domain calculations give the absorbed pump energy, which
drives a four-temperature model coupled to calibrated stochastic Landau--Lifshitz--Bloch dynamics.
A later probe pulse then reads the recovered magnetic state. At the reference fluence, a single pump
pulse reverses \ToggleSwitchPercent{} of the film from either saturated state, and this fraction
changes little over a \FluenceWindowFactor{}-fold range of incident fluence. The film switches
wherever the local absorbed fluence exceeds a single threshold, and applying this threshold to the
calculated absorption map reproduces the switched fraction of the full simulation. When the film starts from zero
magnetization, thermal noise sets the direction of each recovered local moment at random. The
transmitted, reflected, and absorbed fractions of the probe energy differ by at most 0.02 between
the saturated and stochastic states. The magnetic Faraday and Kerr rotations separate these states much
more clearly and are about \FaradaySuppression{} and \KerrSuppression{} times smaller,
respectively, in the stochastic state. The reflected signal also contains a magnetization-independent
geometric background, which can be removed by subtracting a reference signal. The model therefore
predicts that a guided probe can distinguish the written states through its polarization. These
predictions assume a homogeneous film and neglect lateral heat flow and exchange coupling between
neighbouring grid cells.

\vspace{0.9em}
\end{abstract}

\section{Introduction}
\label{sec:intro}

\begin{brief}
Ultrafast optical control of magnetism began with the observation of sub-picosecond
demagnetization in nickel \cite{Beaurepaire1996,Kirilyuk2010}. All-optical switching of the
magnetization was later demonstrated in rare-earth--transition-metal ferrimagnets such as GdFeCo
\cite{Stanciu2007,Radu2011}. In these alloys, rapid heating alone can reverse the magnetization,
independent of the light helicity, because the two antiferromagnetically coupled sublattices
respond on different time scales \cite{Ostler2012,ElHadri2016,Davies2022}. Helicity-dependent
switching has since been reported in ferromagnetic films and engineered multilayers
\cite{Lambert2014,Mangin2014}, and helicity-independent toggle switching in a gadolinium-free
ferrimagnet \cite{Banerjee2020}. Localized-spin models describe these ultrafast dynamics
\cite{Ostler2011,Atxitia2010}.
\end{brief}
\begin{expository}
Ultrafast optical control of magnetism began with the observation of sub-picosecond
demagnetization in nickel \cite{Beaurepaire1996,Kirilyuk2010}. All-optical switching of the
magnetization was later demonstrated in rare-earth--transition-metal ferrimagnets such as GdFeCo
\cite{Stanciu2007,Radu2011}. In these alloys, the transition-metal and rare-earth sublattices are
antiferromagnetically coupled but respond to heating on different time scales. Rapid heating can
therefore produce a transient ferromagnetic-like state and reverse the magnetization independently
of the light helicity \cite{Ostler2012,ElHadri2016,Davies2022}. Helicity-dependent switching has
since been reported in ferromagnetic films and engineered multilayers \cite{Lambert2014,Mangin2014},
and helicity-independent toggle switching in a gadolinium-free ferrimagnet \cite{Banerjee2020}.
Localized-spin models provide an established framework for describing these ultrafast dynamics
\cite{Ostler2011,Atxitia2010}.
\end{expository}
Reviews have covered the materials and device structures for helicity-dependent switching
\cite{ElHadri2017JPhysD} and helicity-independent switching in ferromagnets
\cite{Polley2025JPhysD}, and single-shot switching has also been reported in TbFe
\cite{Jiang2024JPhysD}.

All-optical switching is therefore a candidate for magnetic memory integrated with photonic
circuits, in which guided light would both write a magnetic state and read it. A magnetic element switched in this way needs only
the energy of the optical pulse. It therefore differs from electro-optic modulators, which need an
electrical drive \cite{Wolf2018}, and from high-$Q$ photonic-crystal cavities and all-optical logic
gates, which rely on strong resonant confinement \cite{Akahane2003,Anagha2022}. For the writing
step, guided femtosecond pulses have recently switched a Co/Gd Hall cross patterned on a silicon
nitride waveguide, with an electrically measured contrast of up to 90\% in a 500 nm device
\cite{Li2025onchip}. In larger devices, the contrast decreased and the switching became less
reproducible. The authors attributed this mainly to non-uniform absorption of the guided mode,
with possible contributions from domain-wall relaxation and thermally assisted depinning. Co/Gd
synthetic ferrimagnets have also been switched with picosecond pulses, a duration suited to
integrated photonics \cite{Li2025picosecond}. These studies suggest that the spatial profile of the
guided field is an important factor in the performance of an integrated switching element.

The readout step has developed separately. Integrated magneto-optics already provides isolators,
non-reciprocal circuits, and reconfigurable or in-memory photonic functions
\cite{Srinivasan2022,Wei2025,Kimel2022JPhysD,Pintus2025}, and phase-change memory cells on
waveguides have been written and read by the same guided light \cite{Rios2015}. In ultrafast
magnetic devices, however, writing and readout are usually handled by different means. All-optical
switching has been combined with electrical spintronic detection \cite{Lalieu2019}. In
heat-assisted magnetic recording, light only heats the medium, and a magnetic field writes the bit
that an electrical sensor then reads \cite{Challener2009}. For optical readout, an asymmetric
waveguide that converts Kerr rotation into an intensity difference has been analysed
\cite{Demirer2022}. A hybrid plasmonic--photonic design has also been proposed for switching and
reading a 100 nm magnetic bit \cite{Pezeshki2023}.

\begin{brief}
Modelling the complete sequence involves connected optical, thermal, and magnetic calculations.
Switching models commonly prescribe the temperature history and do not calculate the guided field
\cite{Ostler2012,Atxitia2012,Mentink2012,Evans2014atomistic}. Integrated photonic studies generally
calculate the guided field and its magneto-optic response for a prescribed magnetic state
\cite{Demirer2022,Pezeshki2023,Wei2025}. A spatially uniform model also cannot capture
the position-dependent fluence produced by a confined mode. To our knowledge, no study has yet followed
the complete sequence from spatially resolved pump absorption through magnetic recovery to a later
probe signal. In this work, we ask which properties of a guided probe can distinguish the magnetic
state that an earlier guided pump has written.
\end{brief}
\begin{expository}
Modelling the complete sequence involves connected optical, thermal, and magnetic calculations.
Atomistic spin dynamics, multisublattice Landau--Lifshitz--Bloch models, and phenomenological rate
models commonly prescribe the temperature history
\cite{Ostler2012,Atxitia2012,Mentink2012,Evans2014atomistic}. They do not determine the local
energy deposited by a guided field or the polarization of a later probe. Integrated photonic
studies generally calculate the guided field and its magneto-optic response for a magnetic state that
is given as an input \cite{Demirer2022,Pezeshki2023,Wei2025}. A spatially uniform approximation is
also unsuitable when a confined mode produces a strongly position-dependent absorbed fluence. To
our knowledge, no study has yet followed the complete sequence from spatially resolved pump
absorption through magnetic recovery to a later probe signal. In this work, we ask which properties
of a guided probe can distinguish the magnetic state that an earlier guided pump has written.
\end{expository}

To answer this question, we treat the pump absorption, heating, magnetic dynamics, and probe
response as linked stages of one calculation. We apply this approach to an 8 nm GdFeCo film
inserted across a $60\,\mu\mathrm{m}$ Si$_3$N$_4$ waveguide. A three-dimensional dispersive
finite-difference time-domain solver calculates the pump field and the local absorbed energy. This
energy drives a four-temperature model and calibrated stochastic two-sublattice
Landau--Lifshitz--Bloch dynamics, so the heat source and the temperature history are both outputs
of the calculation. Magneto-optic coupling is disabled during the pump stage. After the film has
relaxed, a separate probe pulse propagates through the fixed magnetic state, which enters the
optical calculation through the gyrotropic permittivity \cite{Inoue2013magnetophotonics}.

The calculations show that non-uniform absorption of the guided mode limits the switched fraction,
and that a threshold rule applied to the absorption map reproduces the result of the full
three-dimensional calculation. When the film starts from zero magnetization, thermal noise selects
the direction of each recovered local moment. For readout, the magnetic Faraday and Kerr rotations
separate the saturated and stochastic states far more clearly than the probe energy fractions,
which change by at most 0.02. The reflected signal also carries a geometric background, which a
reference measurement can separate from the magnetic signal. These optical signatures have yet to
be measured. The calculations also consider only a bare waveguide, and resonant confinement may
increase the absorption and lower the switching threshold \cite{Jiang2025photoniccrystal}.

Section~\ref{sec:model} describes the model. Section~\ref{sec:results} presents and discusses the
results, and section~\ref{sec:convergence} reports the numerical tests. Section~\ref{sec:conclusion}
summarizes the findings.

\section{Physical and numerical model}
\label{sec:model}

\subsection{Device concept and guided excitation}

\begin{figure}[!htb]
  \centering
  \includegraphics[width=0.82\textwidth]{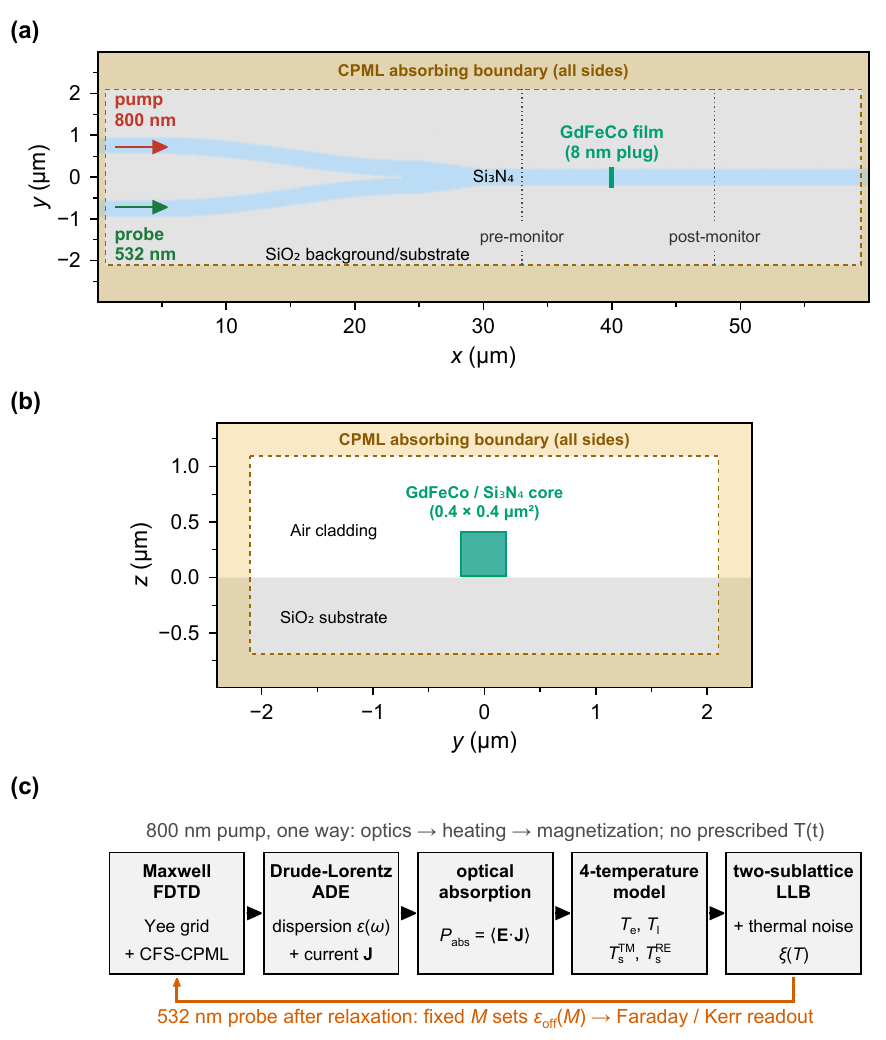}
  \caption{Integrated device and sequential pump--probe model.
    (a)~Waveguide layout with the two inputs, GdFeCo film, monitors, and CPML boundary.
    (b)~Waveguide cross-section. (c)~Optical, thermal, and magnetic calculation stages.}
  \label{fig:device}
\end{figure}

Figure~\ref{fig:device}(a,b) shows the simulated silicon nitride photonic circuit. The waveguide
has a $0.4\times0.4\,\mu\mathrm{m}^2$ cross-section on a SiO$_2$ substrate, with air above it. Two
input arms carry the 800 nm pump and the 532 nm probe and merge into a single output waveguide. An
8 nm GdFeCo film crosses the merged waveguide at $x = 40\,\mu\mathrm{m}$, so both pulses reach the
film as confined guided modes.

The pump wavelength of 800 nm matches that of single-pulse switching experiments on GdFeCo
\cite{Stanciu2007,Radu2011,Ostler2012}. The probe wavelength of 532 nm lies in the visible range,
where silicon nitride guides light with low loss \cite{Buzaverov2024SiN,Bose2024SiN}. It is not
optimized for the magneto-optic response of GdFeCo.

The two pulses are applied in sequence. The film first absorbs part of the pump energy, which
drives ultrafast heating and helicity-independent switching. After the magnetic state has relaxed,
a separate probe pulse propagates through the fixed state. The local magnetization changes the
gyrotropic permittivity of the film, which rotates the polarization of the transmitted field
(Faraday rotation) and of the reflected field (Kerr rotation). These signals are recorded at two
monitor planes, one after the film and one before it [figure~\ref{fig:device}(a)]. The calculation
gives the absorbed fluence, the four reservoir temperatures, the FeCo and Gd sublattice
magnetizations, the local magnetic state, the optical energy fractions, and the polarization
rotations. Supplementary S8 lists the sources and status of the main model inputs.

\subsection{Non-uniform spatial discretization and time-step stability}
\label{sec:mesh}


\begin{brief}
The fields are discretized on a staggered Yee mesh \cite{Yee1966}. Because the geometry combines an
8 nm film, a $0.4\,\mu\mathrm{m}$ core, and a $60\,\mu\mathrm{m}$ circuit, a uniform 4 nm mesh
would require $\approx1.7\times10^{10}$ cells. The graded $4099\times190\times100$ mesh used here
has $\approx7.8\times10^{7}$ cells. Its cell size is 4 nm across the film and 20 nm across the core.
Outside these regions, the cell size grows geometrically as
\end{brief}

\begin{expository}
The fields are discretized on a staggered Yee mesh \cite{Yee1966}. The geometry combines an 8 nm
film, a $0.4\,\mu\mathrm{m}$ core, and a $60\,\mu\mathrm{m}$ circuit. A uniform 4 nm mesh would
therefore require $\approx1.7\times10^{10}$ cells. The graded mesh shown in \Figmesh{} uses
$4099\times190\times100$ cells ($\approx7.8\times10^{7}$), more than two orders of magnitude
fewer. It retains 4 nm cells across the film and a 200 nm surrounding buffer along $x$, and 20 nm
cells across the core along $y$ and $z$. Outside these fine regions, the cell size increases as
\end{expository}
\begin{equation}
  \Delta_{k+1} = \min\!\left(r\,\Delta_k,\;\Delta_{\max}\right), \qquad r = 1.05,
  \label{eq:grading}
\end{equation}
until it reaches $\Delta_{\max}=15$ nm along $x$ or the outer boundary of the domain along $y$ and
$z$. Limiting the growth to 5\% per cell keeps adjacent cells similar in size, which preserves the
local second-order accuracy of the staggered scheme on a smoothly graded mesh
\cite{Taflove2005,JimenezMejia2015}.
\begin{expository}
The coarsest 15 nm propagation cell gives at least 17 points per wavelength for the 532 nm probe in
Si$_3$N$_4$, increasing to about 65 near the film. Larger
transverse cells occur only in the cladding and absorbing boundary, where the guided field is
evanescent.
\end{expository}
We refer to this mesh as the production mesh, and to a full three-dimensional calculation on it as
a production calculation. On this mesh, the film contains \ActiveCellCount{} active magnetic cells.

The time step satisfies the Courant--Friedrichs--Lewy stability condition \cite{Courant1928},
\begin{equation}
  \Delta t = \frac{S}{c_0\sqrt{\Delta_{x,\min}^{-2} + \Delta_{y,\min}^{-2} +
  \Delta_{z,\min}^{-2}}}, \qquad S = 0.99,
  \label{eq:cfl}
\end{equation}
where $S$ is the Courant number and each $\Delta_{\min}$ is the smallest of the primary and
half-offset spacings of the staggered mesh along that axis. The minimum spacings along $x$, $y$, and
$z$ are 4, 6.8, and 20 nm, which give $\Delta t=1.12\times10^{-17}$ s. The $y$ value is the
half-offset spacing of the 13.6 nm cell that closes the mesh at the domain boundary.
\begin{expository}
Section~\ref{sec:convergence} reports the grid-convergence and long-time stability tests.
\end{expository}
\begin{brief}
Section~\ref{sec:convergence} reports the grid-convergence and stability tests.
\end{brief}

\subsection{Dispersive Maxwell solver and absorbed optical work}
\label{sec:maxwell}

The electromagnetic fields obey Maxwell's curl equations, in which polarization currents represent
the dispersive response of the GdFeCo film,
\begin{equation}
  \mu_0\,\partial_t\mathbf{H} = -\nabla\times\mathbf{E}, \qquad
  \varepsilon_0\varepsilon_r(\mathbf{r})\,\partial_t\mathbf{E} =
  \nabla\times\mathbf{H} - \mathbf{J}_P - \mathbf{J}_{\mathrm{MO}}.
  \label{eq:maxwell}
\end{equation}
Here, $\mathbf{J}_P$ is the dispersive current of the film and $\mathbf{J}_{\mathrm{MO}}$ is the
magneto-optic current of section~\ref{sec:coupling}. The equations are advanced with the staggered
Yee update on the mesh of section~\ref{sec:mesh}. The dielectric regions use wavelength-dependent
refractive-index data for Si$_3$N$_4$ and fused silica \cite{RefractiveIndex_Si3N4_Luke,Malitson1965},
and air has a refractive index of one. In the film, each Lorentz--Drude pole $\ell$ is implemented
through an auxiliary differential equation,
\begin{equation}
  \partial_t^2\mathbf{P}_\ell + \gamma_\ell\,\partial_t\mathbf{P}_\ell + \omega_\ell^2\,\mathbf{P}_\ell
  = \varepsilon_0 S_\ell\,\mathbf{E}, \qquad
  \mathbf{J}_P = \sum_\ell \partial_t\mathbf{P}_\ell,
  \label{eq:ade}
\end{equation}
which gives the susceptibility
$\chi_\ell(\omega)=S_\ell/(\omega_\ell^2-\omega^2-\mathrm{i}\gamma_\ell\omega)$. The pole strengths $S_\ell$ are
fitted to the discrete-time response of the update. At 800 nm, the fitted response reproduces the
target permittivity $\varepsilon_{\mathrm{film}}=-1.66+26.1\mathrm{i}$ to within 1\%. This
empirical target is taken from the archived parameter set \cite{MagnetoPhotonicData} and is not
measured in the present work.

Convolutional perfectly matched layers (CPML) terminate the domain \cite{Roden2000}. They are 40
cells thick along the propagation direction and 12 cells thick in the transverse directions
[figure~\ref{fig:device}(a,b)]. For each coordinate $u$, the derivative in \eqref{eq:maxwell} is
replaced by $\partial_u/\kappa_u+\psi_u$, and the memory variable $\psi_u$ is updated as
\begin{equation}
  \psi_u \leftarrow b_u\,\psi_u + a_u\,\partial_u F, \qquad
  b_u = \exp\!\left[-\left(\frac{\sigma_u}{\kappa_u} +
  \alpha_u\right)\frac{\Delta t}{\varepsilon_0}\right].
  \label{eq:cpml}
\end{equation}
Here, $F$ is the differentiated field component. The profiles of $\sigma_u$, $\kappa_u$, and
$\alpha_u$ are cubic and are chosen for a target reflection of $10^{-8}$ at normal incidence.

The absorbed power follows from the local work done by the dispersive currents. Only the positive
part is counted,
\begin{equation}
  P_{\mathrm{abs}}(\mathbf{r},t) = \max\!\left(\mathbf{E}\cdot\mathbf{J}_P,\,0\right), \qquad
  U_{\mathrm{abs}}(\mathbf{r}) = \int P_{\mathrm{abs}}(\mathbf{r},t)\,\mathrm{d}t,
  \label{eq:pabs}
\end{equation}
and $U_{\mathrm{abs}}$ is accumulated separately in every film cell. The spatial profile of the heat
source therefore follows the calculated guided-mode field.
\begin{expository}
The production calculation uses Float64 arithmetic for accumulated quantities and Float32 storage
for the fields and CPML variables, keeping the GPU memory requirement below 6 GB.
\end{expository}

\FloatBarrier
\subsection{Four-temperature and stochastic two-sublattice dynamics}
\label{sec:llb}

The absorbed power drives electron, lattice, FeCo-spin, and Gd-spin reservoirs through a
four-temperature model \cite{Anisimov1974,Koopmans2010}. With
$i\in\{\mathrm{TM},\mathrm{RE}\}$ denoting the FeCo and Gd sublattices,
\begin{equation}
\begin{aligned}
  C_e(T_e)\,\dot T_e &= -G_{el}\,(T_e - T_l)
    - \sum_i G_{es,i}\,(T_e - T_{s,i}) + \bar P_{\mathrm{abs}},
    \qquad C_e = \gamma_e T_e,\\
  C_l\,\dot T_l &= G_{el}\,(T_e - T_l) - C_l\,(T_l - T_0)/\tau_{\mathrm{sub}},\\
  C_{s,i}\,\dot T_{s,i} &= G_{es,i}\,(T_e - T_{s,i}) - Q_{s,i},
    \qquad Q_{s,i} = M_{s,i}\,H^{\mathrm{ani}}_{i,x}\,\partial_t m_{i,x},
\end{aligned}
\label{eq:4tm}
\end{equation}
where $C_e$, $C_l$, and $C_{s,i}$ are the electron, lattice, and spin heat capacities, and $G_{el}$
and $G_{es,i}$ are the electron--lattice and electron--spin coupling constants. The source term
$\bar P_{\mathrm{abs}}$ is the window-averaged optical power of section~\ref{sec:coupling}, and
$T_0$ is the ambient temperature. The lattice cools to the substrate with a time constant
$\tau_{\mathrm{sub}}=2$ ps, and the terms $Q_{s,i}$ account for energy exchanged with the
anisotropy field $H^{\mathrm{ani}}_{i,x}$. Equation~\eqref{eq:4tm} has no lateral heat-flow term,
so each film cell is heated independently. The element-dependent electron--spin coupling represents
the slower response of the rare-earth sublattice measured and modelled in ferrimagnetic alloys
\cite{Koopmans2010,Mekonnen2013,Zhang2022}. In the production calculations, the coupling constants
follow from $G_{es,i}=C_{s,i}/\tau_i$ with $\tau_{\mathrm{TM}}=100$ fs and
$\tau_{\mathrm{RE}}=430$ fs, the element-resolved demagnetization times measured for Fe and Gd in
GdFeCo \cite{Radu2011}. The couplings are thus chosen to reproduce these measured times and are
kept fixed.
\begin{expository}
For numerical stability, the electron equation is integrated using the energy variable
$u_e=\tfrac{1}{2}\gamma_eT_e^2$.
\end{expository}

The reduced magnetization $\mathbf{m}_i$ follows a Landau--Lifshitz--Bloch equation
\cite{Garanin1997,Kazantseva2008,Atxitia2012,Atxitia2017},
\begin{equation}
  \partial_t\mathbf{m}_i = -\gamma\,\mathbf{m}_i\times\mathbf{H}_i
  + \frac{\gamma\alpha_{\parallel,i}}{m_i^2}\,(\mathbf{m}_i\cdot\mathbf{H}_i)\,\mathbf{m}_i
  - \frac{\gamma\alpha_{\perp,i}}{m_i^2}\,
    \mathbf{m}_i\times(\mathbf{m}_i\times\mathbf{H}_i)
  + \Lambda_i\,\mathbf{m}_i + \mathbf{R}_i.
  \label{eq:llb}
\end{equation}
Here, $\gamma$ is the gyromagnetic ratio. Below the Curie temperature, the longitudinal and
transverse damping coefficients are $\alpha_{\parallel,i}=2\alpha_{0,i}T_{s,i}/3T_{\mathrm{C}}$ and
$\alpha_{\perp,i}=\alpha_{0,i}(1-T_{s,i}/3T_{\mathrm{C}})$. The longitudinal term relaxes the
magnetization magnitude towards its equilibrium value $m_{\mathrm{eq},i}$, with
$\Lambda_i\propto\gamma(m_{\mathrm{eq},i}-m_i)/(2\tilde\chi_{\parallel,i}m_{\mathrm{eq},i})$, where
$\tilde\chi_{\parallel,i}$ is the reduced longitudinal susceptibility. The relaxation rate is
limited by the times $\tau_i$ during demagnetization and by calibrated recovery times of 2.8 ps for
FeCo and 3.2 ps for Gd. The effective field contains the intersublattice exchange, uniaxial
anisotropy, and thermal noise,
\begin{equation}
  \mathbf{H}_i = \frac{z_c J_{ij}}{\mu_i}\,\eta(T_e)\,\mathbf{m}_j
  + \frac{2K_i}{M_{s,i}}\,m_{i,x}\,\hat{\mathbf{x}} + \boldsymbol{\xi}_i(t),
  \label{eq:heff}
\end{equation}
where $j$ denotes the other sublattice, $z_c=12$ is the number of nearest neighbours, and the
intersublattice exchange constant is negative, $J_{\mathrm{TM,RE}}<0$. The quantities $\mu_i$,
$M_{s,i}$, $K_i$, and $\alpha_{0,i}$ are the atomic moment, saturation magnetization, anisotropy
constant, and damping parameter of sublattice $i$. The values $\mu_{\mathrm{TM}}=1.92\,\mu_{\mathrm{B}}$,
$\mu_{\mathrm{RE}}=7.63\,\mu_{\mathrm{B}}$, and $\alpha_{0,i}=0.05$ are taken from the atomistic
GdFeCo model of Radu et al \cite{Radu2011}. In the calibrated model, the factor $\eta(T_e)$
weakens the exchange as the electron temperature rises and smoothly decouples the sublattices above
about $0.92T_{\mathrm{C}}$. The equilibrium magnitudes follow from the coupled mean-field Brillouin
equations
\begin{equation}
  m_{\mathrm{eq},i} = B_{S_i}\!\left(\frac{z_c\left[J_{ii}\,m_{\mathrm{eq},i} + J_{ij}\,m_{\mathrm{eq},j}\right]}
       {k_B\,T_{s,i}}\right),
  \label{eq:meq}
\end{equation}
where $B_S$ is the Brillouin function for spin $S$, with $S_{\mathrm{TM}}=1$ and
$S_{\mathrm{RE}}=7/2$.

Thermal fluctuations enter through a Langevin field that is white in space and time
\cite{Brown1963}, with mean and covariance
\begin{equation}
  \langle \xi_{i,a}(t)\rangle = 0, \qquad
  \langle \xi_{i,a}(t)\,\xi_{j,b}(t')\rangle =
  \frac{2\alpha_{0,i}\,k_B\,T_{s,i}}{\gamma\,M_{s,i}\,V_{\mathrm{cell}}}\,
  \delta_{ij}\,\delta_{ab}\,\delta(t - t'),
  \label{eq:fdt}
\end{equation}
where $V_{\mathrm{cell}}$ is the cell volume. At each time step, the field is drawn from
independent Gaussian samples with standard deviation
\[
  \sigma_i = \left(\frac{2\alpha_{0,i}k_BT_{s,i}}
  {\gamma M_{s,i}V_{\mathrm{cell}}\Delta t}\right)^{1/2}.
\]
For the $4\times20\times20\,\mathrm{nm}^3$ film cells at 300 K and the 1 fs relaxation step, this
gives $\sigma_{\mathrm{TM}}\approx\NoiseSigmaTM{}$ and $\sigma_{\mathrm{RE}}\approx\NoiseSigmaRE{}$. In a test at
$1.02T_{\mathrm{C}}$, the mean FeCo magnetization magnitude with noise agrees with that of an
otherwise identical noise-free calculation to within \NoiseNormDeviation{}. A full stochastic Landau--Lifshitz--Bloch treatment has
separate longitudinal and transverse noise terms \cite{Evans2012}. The isotropic field of
\eqref{eq:fdt} is a simplified form, and the test above checks it near $T_{\mathrm{C}}$.

The term $\mathbf{R}_i$ is a phenomenological switching channel based on the exchange-relaxation
picture of multisublattice magnets \cite{Mentink2012}. It relaxes each sublattice towards a target
along the easy axis,
\begin{equation}
  \mathbf{R}_i = \nu_i(T_e, T_{s,i})\,\bigl(\mathbf{m}^{*}_i - \mathbf{m}_i\bigr),
  \qquad \mathbf{m}^{*}_i \parallel s^{*}\hat{\mathbf{x}},
  \label{eq:channel}
\end{equation}
where $s^{*}=\pm1$ is the target sign. The rates $\nu_i$ and the target magnitudes are calibrated
once and then fixed. The calibration reproduces the transient ferromagnetic-like switching sequence
reported for GdFeCo \cite{Radu2011,Ostler2012,Davies2020}. A forced baseline, used as a numerical
reference, sets $s^{*}=-1$ in every cell. The calibrated model sets the target sign to
$\operatorname{sgn}(m_{\mathrm{RE},x})$ when the following conditions are first satisfied:
\begin{equation}
  T_e > T_{\mathrm{C}}, \qquad
  m_{\mathrm{TM}} < 0.25, \qquad
  |m_{\mathrm{RE},x}| > 0.30, \qquad
  m_{\mathrm{TM}} < 0.35\,|m_{\mathrm{RE},x}|.
  \label{eq:latch}
\end{equation}
These conditions require a strongly quenched FeCo moment and a surviving Gd orientation. We call
them the latch criterion, because the target sign is fixed at the first instant they are met.
Before that instant, $\nu_i=0$. The latch-decided fraction is the fraction of cells in which the
criterion is met. When the film starts from zero magnetization, no Gd orientation survives, so the
criterion is not met and the Langevin field alone selects the recovered orientation.

\subsection{Coupling and production protocol}
\label{sec:coupling}

The complete calculation consists of a pump--relaxation stage followed by a separate probe stage,
\begin{equation}
\begin{aligned}
  (\mathbf{E}_{\mathrm{pump}},\mathbf{H}_{\mathrm{pump}})
  &\xrightarrow{\;\eqref{eq:ade},\,\eqref{eq:pabs}\;}
  \bar P_{\mathrm{abs}}
  \xrightarrow{\;\eqref{eq:4tm}\;}
  \mathbf{T}
  \xrightarrow{\;\eqref{eq:llb}\text{--}\eqref{eq:latch}\;}
  \mathbf{m}_{\mathrm{rec}},\\
  \mathbf{m}_{\mathrm{rec}}
  &\xrightarrow{\;\eqref{eq:mo}\;}
  \boldsymbol{\varepsilon}(\mathbf{m}_{\mathrm{rec}})
  \longrightarrow
  (\mathbf{E}_{\mathrm{probe}},\mathbf{H}_{\mathrm{probe}})
  \longrightarrow
  (T,R,A,\theta_{\mathrm{F}},\theta_{\mathrm{K}}).
\end{aligned}
\label{eq:couplingmap}
\end{equation}
During the pump, the calculated optical work is the heat source for \eqref{eq:4tm}, and the
resulting temperatures drive \eqref{eq:llb}--\eqref{eq:latch}. The temperatures set the damping,
equilibrium magnetization, exchange prefactor, noise amplitude, and calibrated switching rates.
During the probe stage, the recovered magnetization is held fixed and changes only the optical
response.

In the probe calculation, the magnetization enters through antisymmetric off-diagonal elements of
the film permittivity,
\begin{equation}
  \varepsilon_{yz}(\omega) = -\varepsilon_{zy}(\omega)
  = \mathrm{i}\,g(\mathbf{m})\,\varepsilon_{\mathrm{film}}(\omega), \qquad
  g(\mathbf{m}) = Q_{\mathrm{TM}}\,m_{\mathrm{TM},x} + Q_{\mathrm{RE}}\,m_{\mathrm{RE},x}.
  \label{eq:mo}
\end{equation}
The antisymmetric form follows the standard magnetophotonic convention
\cite{Inoue2013magnetophotonics}. The weights $Q_{\mathrm{TM}}=0.020$ and $Q_{\mathrm{RE}}=0.006$
are representative phenomenological values scaled to the sublattice moments. They are not measured
independently for the simulated film. The magneto-optic current $\mathbf{J}_{\mathrm{MO}}$ uses the
same pole representation as \eqref{eq:ade}. A Crank--Nicolson midpoint update couples $E_y$ and
$E_z$, so the antisymmetric term does no net work on the field. Because
$\mathbf{J}_{\mathrm{MO}}=0$ during the pump, the magnetization does not act back on the pump
field.

Each electromagnetic step advances \eqref{eq:maxwell}--\eqref{eq:cpml} with the time step of
\eqref{eq:cfl} and accumulates the absorbed power. After every $N_{\mathrm{mp}}=8$ electromagnetic
steps, the thermal--magnetic solver advances \eqref{eq:4tm}--\eqref{eq:latch} over the same
interval, using the power averaged over that window. The forced baseline uses $N_{\mathrm{mp}}=1$.
Temperatures, spin work, and magnetization are updated in double precision (Float64). The
Brillouin equations are solved by two fixed-point iterations, starting from a lookup table.

Each production calculation has three phases. A film that starts saturated first undergoes a 5 ps
thermal--magnetic equilibration without optical excitation. This phase is omitted for a film that
starts from zero magnetization, so that no branch is selected before the pump arrives. The pump
then runs for about 0.8 ps (71310 electromagnetic steps). It is an $E_z$-polarized Gaussian pulse
with $\sigma_{\mathrm{pump}}=40$ fs, equivalent to a full width at half maximum of 94 fs. Finally,
the system relaxes for 15.2 ps with a 1 fs time step and no optical field or heating. The
independent stochastic calculations use the random seeds 1001--1016.

\subsection{Probe readout and reference normalization}
\label{sec:probe}

The probe is a Gaussian pulse with $\sigma_{\mathrm{probe}}=24$ fs, equivalent to a full width at
half maximum of 57 fs. The two monitor planes of figure~\ref{fig:device}(a) lie before and after
the film and are called the pre- and post-planes. At each plane, every field component
$F\in\{E_y,E_z,H_y,H_z\}$ is recorded through its discrete Fourier transform, and the net forward
energy flux $W_{\mathcal{P}}$ is integrated,
\begin{equation}
  \hat F(y,z;\omega) = \sum_{n} F^{\,n}(y,z)\,
  \mathrm{e}^{\,\mathrm{i}\omega n\Delta t}\,\Delta t,
  \qquad
  W_{\mathcal{P}} = \int\!\mathrm{d}t \iint_{\mathcal{P}}
  \left(E_y H_z - E_z H_y\right)\mathrm{d}A,
  \label{eq:dft}
\end{equation}
and the spectral cross-power is
$\hat S_x(\omega)=\iint(\hat E_y\hat H_z^*-\hat E_z\hat H_y^*)\,\mathrm{d}A$. A reference
simulation uses the same mesh, source, and monitors but removes the film currents. It gives the
incident energy $W_{\mathrm{inc}}=W_{\mathrm{pre}}^{\mathrm{ref}}$ and the incident spectral power
$\hat S_x^{\mathrm{inc}}=\hat S_{x,\mathrm{pre}}^{\mathrm{ref}}$. Its transmission of 0.999 at the
post-plane confirms that the waveguide without the film is nearly lossless. The energy fractions are
\begin{equation}
  T = \frac{W_{\mathrm{post}}}{W_{\mathrm{inc}}}, \qquad
  R = \frac{W_{\mathrm{inc}} - W_{\mathrm{pre}}}{W_{\mathrm{inc}}}, \qquad
  A = \frac{\sum_{c} U_{\mathrm{abs},c}\,V_c}{W_{\mathrm{inc}}}.
  \label{eq:tra}
\end{equation}
Here, $R$ is the forward energy missing at the pre-plane, and $A$ is calculated directly from the
dispersive work \eqref{eq:pabs}, with $V_c$ the volume of film cell $c$. The sum $T+R+A$ is
therefore an independent energy check. We attribute its small deficit from unity to scattering and
residual field energy.
\begin{expository}
Frequency-dependent transmission and reflection are calculated from the corresponding spectral
cross-powers.
\end{expository}

To evaluate the polarization, the monitored fields are first projected onto the guided-mode
profile $w(y,z)$ at the probe frequency $\omega_0$. The projected Jones amplitudes give the
rotation from the launched $E_z$ polarization,
\begin{equation}
  \mathcal{J}_F = \iint w(y,z)\,\hat F(y,z;\omega_0)\,\mathrm{d}A, \qquad
  \theta = \tfrac{1}{2}\operatorname{atan2}\!\left(2\,\mathrm{Re}\,(\mathcal{J}_{E_y}\mathcal{J}_{E_z}^{*}),\;
    |\mathcal{J}_{E_z}|^2 - |\mathcal{J}_{E_y}|^2\right),
  \label{eq:jones}
\end{equation}
and the ellipticity is
$\chi=\tfrac{1}{2}\arcsin[2\,\mathrm{Im}(\mathcal{J}_{E_y}\mathcal{J}_{E_z}^*)/
(|\mathcal{J}_{E_y}|^2+|\mathcal{J}_{E_z}|^2)]$. Equation~\eqref{eq:jones} gives the Faraday
rotation at the post-plane. At the pre-plane, the incident and reflected fields overlap, so the
reflected Jones vector is obtained by subtracting the reference field,
\begin{equation}
  \mathcal{J}^{\mathrm{refl}}_F = \mathcal{J}^{\mathrm{pre}}_F -
  \mathcal{J}^{\mathrm{inc}}_F,
  \label{eq:kerr}
\end{equation}
and \eqref{eq:jones} is then applied to $\mathcal{J}^{\mathrm{refl}}$ to obtain the Kerr rotation.
The $+M$ and $-M$ states, a uniform zero-magnetization state, and the stochastic state are
simulated with identical geometry and monitors. Their rotations are separated into a
magnetization-independent background, given by the uniform zero-magnetization state, and an
odd-in-$M$ contribution,
\begin{equation}
  \bar\theta = \theta(M{=}0), \qquad
  \theta^{\mathrm{MO}}_{\pm} = \theta(\pm M) - \bar\theta, \qquad
  \tfrac{1}{2}\left[\theta(+M) + \theta(-M)\right] \approx \bar\theta,
  \label{eq:pedestal}
\end{equation}
where the final relation checks the background subtraction. It holds to $0.004^{\circ}$ for the
Faraday rotation and to $0.69^{\circ}$ for the Kerr rotation (Supplementary S6). The saturated
magnetic rotations reported below are therefore the odd part,
$\tfrac{1}{2}|\theta(+M)-\theta(-M)|$, which does not depend on $\bar\theta$.

\subsection{Statistical tests and convergence measures}
\label{sec:verification}

\begin{expository}
Changes to the simulation workflow are checked with a 2000-step comparison of the common output
arrays. One stochastic realization is also repeated to confirm reproducibility for a fixed seed.
\end{expository}
Each cell of a stochastic calculation ends in one of two branches. A cell is in the negative
branch when $m_{\mathrm{TM},x}<0$ and $m_{\mathrm{RE},x}>0$, that is, when the $x$ component of the
FeCo moment is negative and that of the antiparallel Gd moment is positive. In the positive branch,
both signs are reversed. The statistical tests check the null hypothesis that each cell selects
either branch independently and with equal probability. For realization $s=1,\ldots,N_s$, the
final state defines a binary mask over the $N_c=\ActiveCellCount{}$ active cells, with
$b_c^{(s)}=1$ if cell $c$ ends in the negative branch and $b_c^{(s)}=0$ otherwise. The
negative-branch occupancy and its ensemble $z$ score are
\begin{equation}
  \hat P_s = \frac{1}{N_c}\sum_{c} b_c^{(s)}, \qquad
  z = \frac{\langle \hat P\rangle - \tfrac{1}{2}}{\sigma_{\hat P}/\sqrt{N_s}},
  \label{eq:mask}
\end{equation}
where $N_s=\SeedCount{}$, and $\langle \hat P\rangle$ and $\sigma_{\hat P}$ are the mean and
standard deviation of $\hat P_s$ over the realizations. For independent branch selection, the 95\%
interval for one realization is $\tfrac{1}{2}\pm1.96\sqrt{1/(4N_c)}$. For each cell, the number of
realizations that end in the negative branch, $k_c=\sum_s b_c^{(s)}$, is compared with the
binomial distribution $\mathrm{Binomial}(N_s,1/2)$ through
\begin{equation}
  E_k = N_c\binom{N_s}{k}2^{-N_s}, \qquad
  \chi^2 = \sum_{g}\frac{(O_g - E_g)^2}{E_g}.
  \label{eq:chisq}
\end{equation}
Here, $E_k$ is the expected number of cells with $k_c=k$, and $O_g$ and $E_g$ are the observed and
expected counts in bin $g$. Adjacent bins are combined until every expected count satisfies
$E_g\ge5$. Independence between realizations is also tested with the pairwise mask agreement
$a_{ss'}$, the fraction of cells with $b_c^{(s)}=b_c^{(s')}$, over all
$\binom{N_s}{2}=\PairCount{}$ pairs. For independent and unbiased selection, its expected value is
$1/2$.

Electromagnetic convergence is evaluated against a 22 nm reference grid. Benchmark traces
$u_i(t)$ are calculated on seven grids $i$ from 50 to 25 nm, sampled on a common time axis, and
reduced to the root-mean-square error
\begin{equation}
  \varepsilon_i = \left[\left\langle
  (u_i - u_{\mathrm{ref}})^2\right\rangle_t\right]^{1/2}, \qquad
  \varepsilon_i = K\left(\Delta x_i^{\,p} - \Delta x_{\mathrm{ref}}^{\,p}\right).
  \label{eq:l2fit}
\end{equation}
The prefactor $K$ and the convergence order $p$ are obtained by least squares, and the term
$-\Delta x_{\mathrm{ref}}^p$ accounts for the finite error of the reference grid. Long-time
stability is tested with the ratio
\begin{equation}
  \rho_i = \frac{\max_{t \,\in\, \mathrm{last}\ 10\%}\,|e_i(t)|}
                {\max_{t \,\in\, \mathrm{first}\ 90\%}\,|e_i(t)|},
  \label{eq:stability}
\end{equation}
where $e_i(t)=u_i(t)-u_{\mathrm{ref}}(t)$. The ratio is of order unity for a bounded error and
large for an error that grows with time. A grid passes the test when $\rho_i\le1.5$, and the
criterion is applied at every tested resolution. Section~\ref{sec:convergence} reports the dielectric and dispersive
tests.

\subsection{Hybrid fluence mapping}
\label{sec:fluencemap}

The fluence dependence is calculated in two ways, with a dense sweep of a single film cell and
with five full three-dimensional calculations, and the two are compared through the absorbed
fluence. In the single-cell calculation, the noise is switched off and the calibrated model
\eqref{eq:4tm}--\eqref{eq:latch} starts from $+M$. The optical calculation is replaced by a Gaussian
absorbed-power density,
\begin{equation}
  \bar P_{\mathrm{abs}}(t) = p_0\,
  \mathrm{e}^{-\left[(t - t_p)/\tau_p\right]^2},
  \qquad
  F^{\mathrm{0D}}_{\mathrm{abs}} = d\int \bar P_{\mathrm{abs}}\,\mathrm{d}t
  = \sqrt{\pi}\,p_0\,\tau_p\,d,
  \label{eq:fluence0d}
\end{equation}
where $p_0$ is the peak power density, $t_p$ is the time of the peak, $\tau_p=0.2$ ps, and $d=8$ nm
is the film thickness. The 46-point sweep covers 0.06--1.59 $\mathrm{mJ\,cm^{-2}}$. A cell is classified as
switched when $m_{\mathrm{TM},x}<0$ and $m_{\mathrm{RE},x}>0$ in the final state, and the threshold
is the midpoint between the last sample that does not switch and the first sample that does.

The complete three-dimensional calculation is repeated with the pump field scaled by the factors
$s\in\{1.00,1.15,1.27,1.40,1.55\}$. The scale $s=1.27$ is the reference condition, and the five
incident fluences are 0.62--1.49 times its value. The film-averaged absorbed fluence is
\begin{equation}
  F^{\mathrm{3D}}_{\mathrm{abs}}
  = \frac{\sum_c U_{\mathrm{abs},c}\,V_c}{\sum_c V_c/d},
  \label{eq:fluence3d}
\end{equation}
which is the total absorbed energy divided by the effective film area. The five points follow
$F^{\mathrm{3D}}_{\mathrm{abs}}=0.504s^2$ $\mathrm{mJ\,cm^{-2}}$ with relative deviations below
$10^{-7}$, which confirms that the optical absorption is linear. The single-cell and
three-dimensional fluence scales are matched by equating their peak electron temperatures,
$T_e^{\mathrm{peak,0D}}(F)=T_e^{\mathrm{peak,3D}}(s)$. The single-cell fluence is the local energy
deposit that sets the switching threshold, whereas \eqref{eq:fluence3d} averages the deposit over
the non-uniform footprint of the guided mode.
\section{Results and discussion}
\label{sec:results}

\begin{stub}
\subsection{Guided-pump absorption and non-equilibrium heating}

The guided 800 nm pump deposits a film-averaged absorbed fluence of \ProductionFluence{} in the
GdFeCo film. The deposition is non-uniform and follows the intensity profile of the guided mode.
Because the film permittivity is spatially uniform in the model, this variation comes from optical
confinement alone. The absorbed energy enters the electron reservoir first, which reaches \TePeak{}
during the 0.8 ps pump window. The FeCo and Gd spin reservoirs then rise above
$T_{\mathrm{C}}=\CurieTemperature{}$ at different rates, while the lattice stays below
$T_{\mathrm{C}}$. This sequence of fast electron heating followed by slower, element-dependent spin
heating is consistent with measurements of ultrafast demagnetization in elemental magnets and in GdFeCo
\cite{Koopmans2010,Radu2011,Zhang2022}. \SecheatingC{} gives the spatial absorption map and the full reservoir dynamics.
\end{stub}

\FloatBarrier
\subsection{Deterministic switching from saturated states}
\label{sec:deterministic}

\begin{figure}[!htb]
  \centering
  \includegraphics[width=\textwidth]{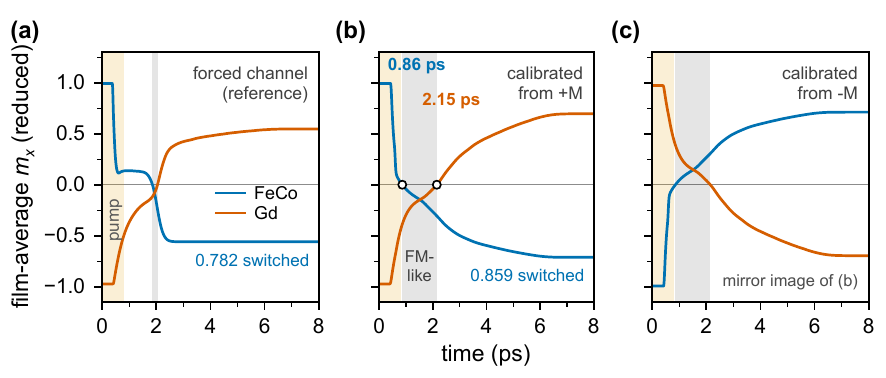}
  \caption{Deterministic FeCo and Gd switching.
    (a)~Forced baseline. (b),(c)~Calibrated switching from $+M$ and $-M$.
    Circles mark zero crossings. Orange and grey bands mark the pump and transient
    ferromagnetic-like intervals.}
  \label{fig:deterministic}
\end{figure}

Figure~\ref{fig:deterministic} shows the response of the FeCo and Gd sublattices to the absorbed
pump energy. Antiferromagnetic exchange keeps the two sublattices oppositely aligned before and
after the pulse. The initial temperature of 300 K lies above the compensation temperature,
$T_{\mathrm{comp}}=\CompensationTemperature{}$, so the FeCo saturation magnetization is larger than
the Gd value and the net magnetization points along the FeCo direction.

The pump drives both spin reservoirs above $T_{\mathrm{C}}$. The magnetization magnitudes of both
sublattices fall towards zero and then recover with the opposite sign as the film cools. The
reversal therefore proceeds through a loss and recovery of magnetic order, with no coherent
rotation of the moments. In the calibrated model, the factor $\eta(T_e)$ weakens the
intersublattice exchange during the hot stage. This temporary weakening allows the two sublattices
to cross zero at different times before their antiparallel alignment is restored.

In figure~\ref{fig:deterministic}(b), FeCo crosses zero at \FeCoCrossing{} and Gd at
\GdCrossing{}. During the \SublatticeDelay{} interval between these crossings, both sublattices
have the same sign, so the film passes through a transient ferromagnetic-like state. The FeCo and
Gd spin temperatures cross $T_{\mathrm{C}}$ only 53 fs apart, so the delay arises mainly from the
magnetic response of the two sublattices. Gd responds more slowly because, compared with FeCo, it
couples more weakly to the electron reservoir, carries a larger atomic moment, and experiences a
smaller exchange field. This order and delay agree with the element-resolved picture of GdFeCo
switching \cite{Radu2011,Mentink2012,Davies2020}.

Panel (a) shows the forced baseline, in which the target direction is fixed. Its switched fraction
is \ForcedSwitchFraction{}. Because the same target would apply to either initial state,
this baseline cannot represent toggling. In the calibrated calculation, the surviving Gd sign
determines the target direction. The switched fraction is then \ToggleSwitchFraction{} from either
saturated state, and panels (b) and (c) are mirror images. This symmetry confirms that the
implementation has no preferred magnetic direction. The latch-decided fraction is
\LatchDecidedFraction{}, and the latched cells include every switched cell. The remaining latched
cells, 1.7\% of the film, return to their initial state, as do the cells that do not satisfy the
criterion.

The final film-averaged magnetization is well below unity in magnitude, although the individual
cells recover to a mean magnitude of \PerCellSaturation{}. The reduction comes from averaging over
switched and unswitched cells. For a switched fraction $f$ and a common local magnitude
$m_{\mathrm{sat}}$,
\begin{equation}
  \bar m_{\mathrm{TM},x}=m_{\mathrm{sat}}(1-2f).
  \label{eq:filmavg}
\end{equation}
With $f=\ToggleSwitchFraction{}$, this gives the computed average of \FilmAverageToggle{}. The
unswitched cells lie mainly at the weakly heated edge of the film, which links the incomplete
reversal to the non-uniform absorption profile. Supplementary S4 lists the sublattice parameters
and gives the checks of the forced baseline and of the film average.

The calibrated model thus produces deterministic toggling and a transient ferromagnetic-like state.
Both results rest on two calibrated rules, the temperature-dependent exchange reduction and the
latch criterion, which are not derived from microscopic theory. The calculation tests the
device-scale consequences and the internal consistency of these rules. Identifying a unique
microscopic switching mechanism is left to first-principles treatments.

\FloatBarrier
\subsection{Stochastic recovery from zero magnetization}
\label{sec:shattering}

The deterministic results of section~\ref{sec:deterministic} start from a saturated film, so the
initial magnetization provides the sign that is reversed. We next start every cell from zero
magnetization. The pump remains helicity-independent, and the magneto-optic term is disabled during
heating, so the optical field acts only as a heat source. The latch criterion is then not met
(section~\ref{sec:llb}), and in all \SeedCount{} realizations thermal noise alone selects the
recovered orientation.

\begin{figure}[!htb]
  \centering
  \includegraphics[width=\textwidth]{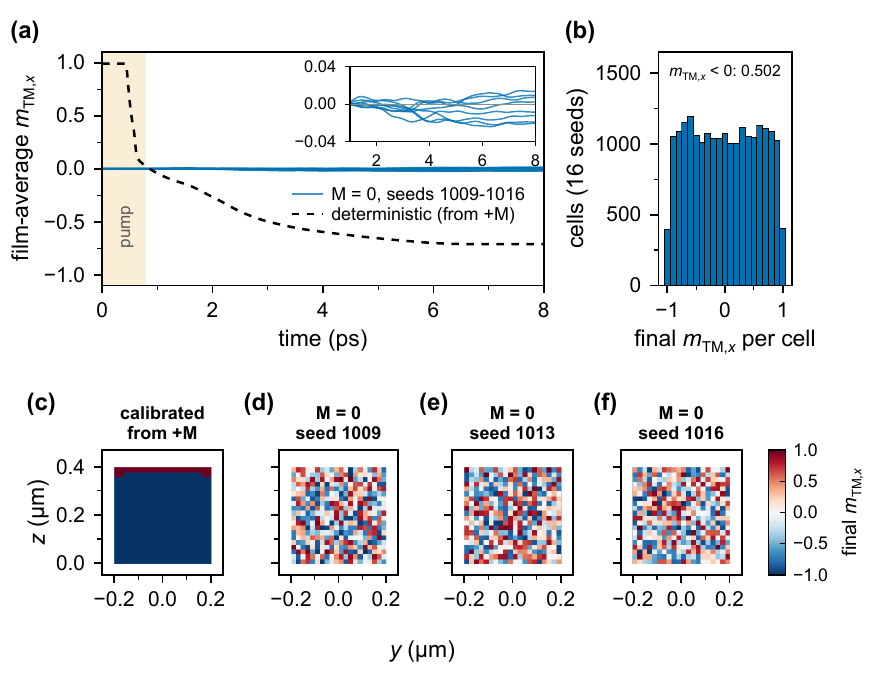}
  \caption{Recovery from zero initial magnetization.
    (a)~Film-averaged FeCo response for eight stochastic realizations and the deterministic
    reference. (b)~Final cell projections for all realizations. (c)~Switched mask of the calibrated
    calculation from $+M$.
    (d)--(f)~Three stochastic branch masks.}
  \label{fig:shattering}
\end{figure}

Figure~\ref{fig:shattering}(a) shows that the film-averaged magnetization of every realization stays
within $\pm0.04$ of zero, and its final value varies by only 0.019 (one standard deviation) between
realizations. The final FeCo projections of the individual cells in panel (b) nevertheless extend
from about $-0.99$ to $+0.99$, with a standard deviation of 0.579. This value is close to
$1/\sqrt{3}=0.577$, as expected for one Cartesian component of unit vectors with random
orientations. The local moments therefore recover to nearly full magnitude in random directions, and
within each cell the FeCo and Gd moments remain nearly antiparallel.

The spatial masks show how the two starting conditions differ. The switched mask of the calibrated
calculation from $+M$ [figure~\ref{fig:shattering}(c)] is a single region that follows the
guided-mode absorption profile, as section~\ref{sec:fluence} shows. The zero-magnetization
calculations give a different pattern for each noise realization [figure~\ref{fig:shattering}(d)--(f)]. Their
nearest-neighbour sign agreement is 0.488--0.513, close to the value of 0.5 expected for
independent choices. Because the model has no exchange stiffness between neighbouring cells, this
cell-scale texture does not predict the physical domain size. The calculation shows stochastic,
spatially disordered recovery on the simulated time scale, but a real film may later form wider
domains through exchange-driven coarsening and domain-wall motion. A micromagnetic treatment with
exchange and domain-wall dynamics could predict that later texture \cite{Raposo2022}. Such later
processes may contribute to the size-dependent contrast and reproducibility reported for on-chip
Co/Gd switching \cite{Li2025onchip}.

Averaged over all realizations, the negative-branch occupancy is \OccupancyMean{} (\Secens{}), and
the fluence profile produces no measurable sign preference. Cells above and below the median local
fluence have occupancies of 0.499 and 0.505. The correlation between the local fluence and the
per-cell count $k_c$ is $r=-0.046$, smaller in magnitude than the 5\% significance threshold of
0.054. The heating profile therefore controls whether a cell reaches the switching regime, and
thermal noise selects its recovered orientation. This is the stochastic state used in the readout
analysis of section~\ref{sec:readout}.

\FloatBarrier
\subsection{Fluence dependence of the switched fraction}
\label{sec:fluence}

\begin{figure}[!htb]
  \centering
  \includegraphics[width=\textwidth]{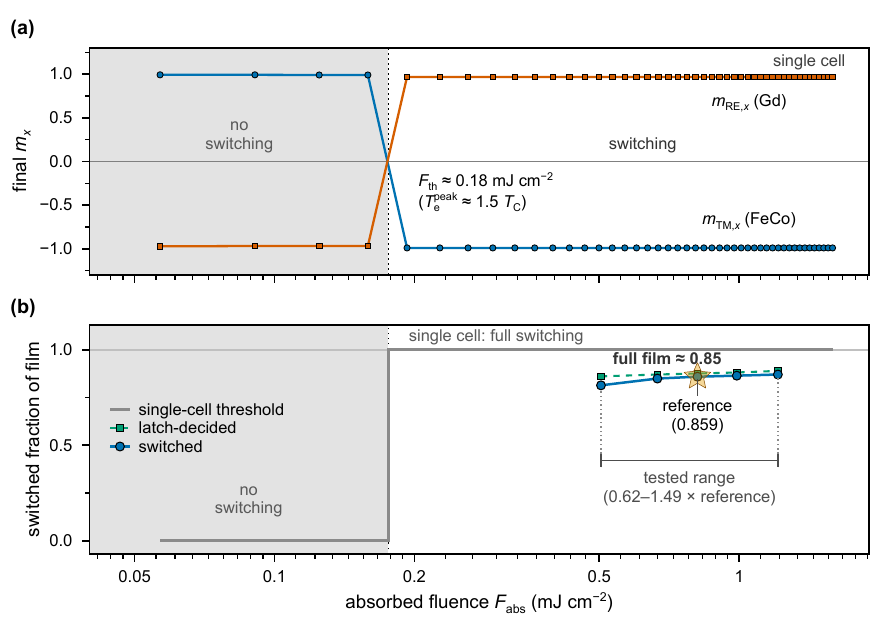}
  \caption{Switching versus fluence.
    (a)~Final single-cell magnetizations. (b)~Local threshold (grey), three-dimensional switched
    fraction (blue), and latch-decided fraction (green). The star marks the reference condition.}
  \label{fig:fluence}
\end{figure}

Figure~\ref{fig:fluence}(a) shows the local switching threshold obtained from the single-cell
calculation. Below $F_{\mathrm{th}}=\FluenceThreshold{}$, the cell returns to its initial state.
Above this threshold, both sublattices reverse and recover close to their full local magnitudes.
The threshold corresponds to a peak electron temperature of about $1.5\,T_{\mathrm{C}}$. The
single-cell calculation gives the absorbed fluence needed to switch one location. The
three-dimensional calculations below add the spatial intensity profile of the waveguide mode.

Figure~\ref{fig:fluence}(b) shows the results of five complete three-dimensional calculations. The
incident fluence spans \FluenceWindowRange{} times the reference value, a
\FluenceWindowFactor{}-fold range. Across this range, the switched fraction changes from 0.814 to
0.871, and the latch-decided fraction from 0.861 to 0.890. At the reference condition, the switched
fraction is \ToggleSwitchFraction{}.

Non-uniform absorption explains why the device shows a broad plateau while a single cell has a
sharp threshold. Each film cell receives a different local fluence and switches when that fluence
exceeds the common threshold. At the reference condition, the largest local absorbed fluence among
unswitched cells is $0.175\,\mathrm{mJ\,cm^{-2}}$, and the smallest among switched cells is
$0.182\,\mathrm{mJ\,cm^{-2}}$. These values bracket $F_{\mathrm{th}}$, so the switched mask in
figure~\ref{fig:shattering}(c) is binary even though the absorbed energy varies continuously. If the
absorption pattern scales linearly with the film-averaged absorbed fluence $\bar F$ of
equation~\eqref{eq:fluence3d}, the switched fraction is
\begin{equation}
  f(\bar F)=\frac{N_{>}(\bar F)}{N_c}.
  \label{eq:cdfbridge}
\end{equation}
Here, $N_{>}(\bar F)$ is the number of cells whose scaled local fluence,
$F_{\mathrm{local},c}\,\bar F/\bar F_{\mathrm{ref}}$, exceeds $F_{\mathrm{th}}$, with
$\bar F_{\mathrm{ref}}$ the value at the reference condition.
This expression combines the single-cell threshold with the distribution of absorbed fluence in the
full device, and it reproduces all five three-dimensional results to within $8\times10^{-4}$.

Close agreement is expected, because the cells are not coupled laterally in the model and each cell
switches according to its own absorbed energy. The comparison therefore tests whether a threshold
found in a noise-free single-cell calculation, with a simplified Gaussian heating pulse, carries
over to the full calculation. A threshold in the fluence absorbed in the magnetic layer has also
been observed experimentally in GdFeCo \cite{Khorsand2012}. In a real film, lateral heat flow and
exchange between cells, both absent here, could blur the threshold contour. In GdFeCo, angular
momentum transfer between nanoscale regions and electronic heat currents are both known to affect
switching \cite{Graves2013,Wilson2017}. Within these assumptions, the result provides a quantitative
basis for the non-uniform-absorption explanation proposed for the size-dependent contrast of on-chip
switching \cite{Li2025onchip}.

The plateau stays below unity because the edge of the film is only weakly illuminated. Increasing
the incident fluence moves the threshold contour further into this low-intensity region, with
diminishing gains in the total switched fraction. This suggests that a more uniform guided-mode
profile could enlarge the switched area more efficiently than a further increase in pulse energy.
We do not test mode-profile optimization here, but related threshold-engineering and
photonic-crystal studies have shown that changing the optical environment can reduce the energy
needed for all-optical switching \cite{Li2021threshold,Jiang2025photoniccrystal}.

The switched and latch-decided fractions do not decrease anywhere in the simulated range, which
extends to $1.59\,\mathrm{mJ\,cm^{-2}}$ in the single-cell sweep and to
$1.21\,\mathrm{mJ\,cm^{-2}}$ in the three-dimensional calculations. Within the model, the upper
boundary of the switching window therefore lies above this range and is not determined here. At higher
fluences, the Gd orientation that sets the toggle target could be lost, and the recovery could
become stochastic. In GdFeCo films, a multidomain state forms at the centre of the spot above a
second, higher fluence threshold \cite{Wei2021}. Experimental switching also depends strongly on
pulse duration and magnetic-stack design \cite{Li2025picosecond,Davies2022,ElHadri2017JPhysD}.
Material damage, interdiffusion, and ablation lie outside the model as well, so the simulated window
is specific to the modelled film and pulse. Supplementary S5 extrapolates the switched fraction to
higher fluence and discusses how the model threshold compares with measured ones.

\FloatBarrier
\subsection{Magneto-optic readout of the magnetic state}
\label{sec:readout}

\begin{figure}[!htb]
  \centering
  \includegraphics[width=0.82\textwidth]{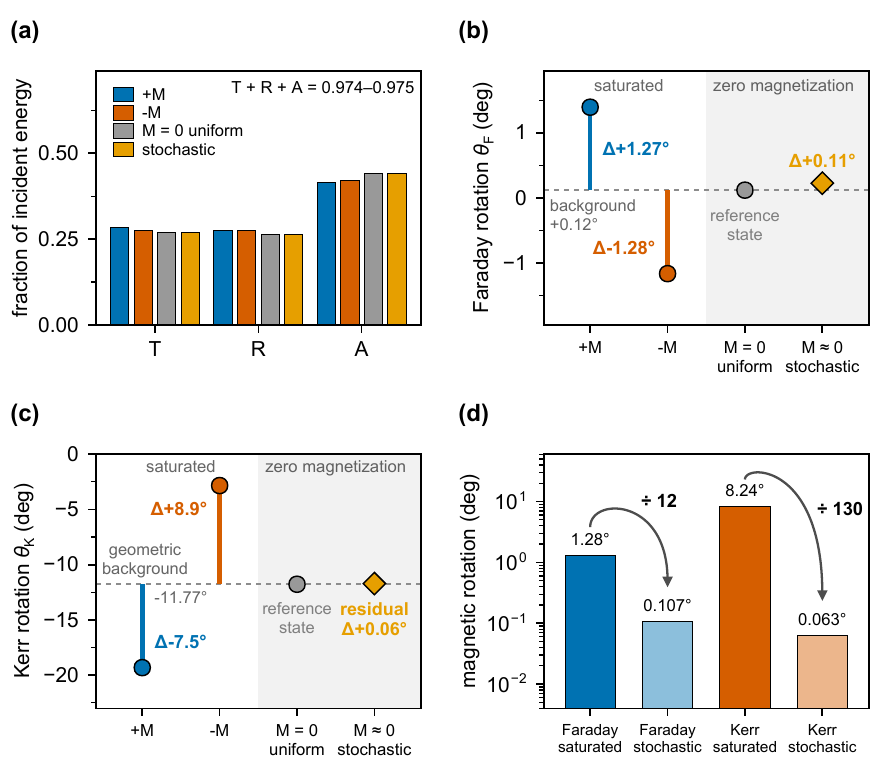}
  \caption{Optical readout of four magnetic states.
    (a)~Energy fractions. (b)~Faraday rotation. (c)~Kerr rotation.
    (d)~Background-subtracted magnetic rotations. Markers show total rotations, and stems show
    magnetic contributions.}
  \label{fig:readout}
\end{figure}

Figure~\ref{fig:readout} compares the response of the 532 nm probe to four magnetic states: the two
saturated states, a uniform zero-magnetization state, and the stochastic state of
section~\ref{sec:shattering}. Each state is held fixed during the probe calculation. The uniform state has no local magnetic
moments and therefore isolates the polarization change produced by the waveguide geometry. The
stochastic state consists of recovered local moments with random orientations.

Figure~\ref{fig:readout}(a) shows that the probe energy fractions depend only weakly on the magnetic
state. Absorption is $0.42$--$0.44$, and transmission and reflection are each about $0.27$--$0.28$.
The strong absorption comes from the metallic optical response of the film at 532 nm. In the probe
calculation, this energy does not act on the fixed magnetic state. Because the three fractions
change by at most 0.02 among the four states, they carry little information about the stored
magnetization. This weak dependence is consistent with the design of integrated readout devices,
which convert the magneto-optic polarization change into an intensity contrast and do not rely on
the total transmitted power \cite{Demirer2022,Pezeshki2023}.

The polarization rotations separate the states more clearly [figure~\ref{fig:readout}(b,c)]. The
uniform zero-magnetization state defines a geometric background of \FaradayPedestal{} in
transmission and \KerrPedestal{} in reflection. After subtracting these backgrounds, the saturated
film gives magnetic rotation magnitudes of \FaradaySatOdd{} for the Faraday channel and
\KerrSatOdd{} for the Kerr channel. The two saturated directions produce opposite signs, as
expected from the odd dependence of the gyrotropic permittivity on magnetization \cite{Sato2022}.

In reflection, the geometric background has a magnitude of $11.77^{\circ}$, about 1.4 times the
average saturated magnetic contribution of \KerrSatOdd{}. A raw Kerr measurement would therefore
contain two contributions of similar size. The numerical analysis separates them using the
zero-magnetization reference and the coherent subtraction of equation~\eqref{eq:kerr}. In an
experiment, an equivalent reference or a differential measurement would serve the same purpose.
Geometry-dependent polarization of this kind is a known concern in integrated magneto-optics
\cite{Wei2025,Kimel2022JPhysD}.

The stochastic state produces much smaller magnetic rotations. Its Faraday residual is
\FaradayMZeroOdd{}, and its Kerr residual is \KerrMZeroResidual{}. Relative to the saturated film,
these values correspond to suppression factors of about \FaradaySuppression{} and
\KerrSuppression{}, respectively [figure~\ref{fig:readout}(d)]. The suppression arises from
cancellation between regions whose magnetization has opposite $x$ components, which
section~\ref{sec:probefields} examines in the probe field. Because the residual depends on the
finite simulated area and on the non-uniform optical weighting of the cells, both suppression
factors are specific to the present model and geometry.

Of the two polarization channels, reflection gives the larger magnetic rotation, but it also carries
the larger geometric background and propagates back towards the input. Whether either signal can be
collected with sufficient stability in a fabricated device remains to be tested experimentally.
Supplementary S6 details the energy accounting and estimates the signal at a detector.

\FloatBarrier
\subsection{Spatial origin of the readout signal}
\label{sec:probefields}

\begin{figure}[!htb]
  \centering
  \ifarxiv
    \includegraphics[width=0.78\textwidth]{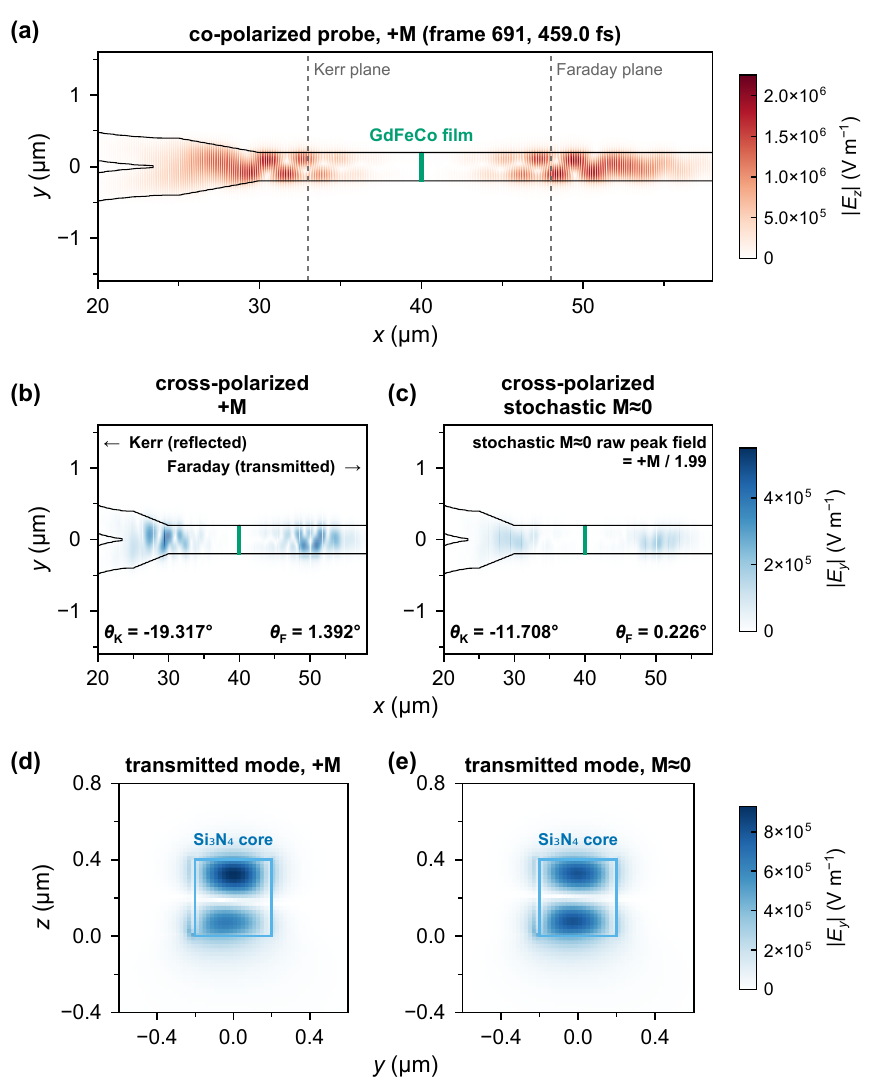}
  \else
    \includegraphics[width=0.66\textwidth]{probe_fields.pdf}
  \fi
  \caption{Probe fields for saturated and stochastic states.
    (a)~Co-polarized field and monitors. (b),(c)~Cross-polarized field maps.
    (d),(e)~Transmitted cross-polarized profiles. Values in (b),(c) are total rotations before
    background subtraction.}
  \label{fig:probefields}
\end{figure}

Figure~\ref{fig:probefields} shows how the magnetic state modifies the guided probe field. The
co-polarized component $E_z$ changes little between magnetic states
[figure~\ref{fig:probefields}(a)], which matches the weak state dependence of the transmission,
reflection, and absorption in figure~\ref{fig:readout}(a). Both the co-polarized field and these
energy fractions are dominated by the diagonal optical response of the film.

The cross-polarized component $E_y$ [figure~\ref{fig:probefields}(b,c)] is generated by the
gyrotropic coupling \eqref{eq:mo}, whose strength is linear in the signed FeCo and Gd
magnetizations. In a saturated film, the local contributions have the same sign across the
illuminated region and add constructively. In the stochastic state, regions with opposite $x$
components contribute with opposite signs and largely cancel. As a result, the peak cross-polarized
amplitude of the saturated state is \XYPeakRatio{} times that of the stochastic state. This ratio is
smaller than the ratios of the background-subtracted rotations, because the field maps also contain
the geometric cross-polarized field, which is present in both states. The rotations annotated in
panels (b) and (c) are likewise total values, $1.39^{\circ}$ in transmission and $-19.3^{\circ}$ in
reflection for the saturated state. Subtracting the geometric backgrounds of
section~\ref{sec:readout}, \FaradayPedestal{} and \KerrPedestal{}, gives the magnetic contributions
shown in figure~\ref{fig:readout}.

The transmitted profiles provide a second signature [figure~\ref{fig:probefields}(d,e)]. Each
profile has a node near the mid-plane of the waveguide and two lobes, the regions of high amplitude
above and below this node. The integrated cross-polarized powers of the two states agree to within \TransmittedParity{}, but the
states divide this power differently between the two lobes. The ratio of the upper to the lower lobe
is \LobeBalancePlus{} for the saturated state and \LobeBalanceZero{}, close to balance, for the
stochastic state. The geometric part of the cross-polarized field is nearly symmetric about the
mid-plane of the waveguide, and the magnetic contribution breaks this symmetry. When the signed
magnetic contributions cancel, the symmetric geometric part dominates.

The field maps show that the predicted readout contrast is distributed across the whole guided
mode. They also suggest that the mode shape could serve as a readout observable alongside the
polarization rotation. Such a readout would extend the integrated-readout strategy of converting
magneto-optic polarization changes into spatial or intensity differences \cite{Demirer2022}. A
mode-selective detector and its fabrication tolerance are not modelled here. Supplementary S7
examines the origin of the lobe imbalance further.

\FloatBarrier
\subsection{Numerical verification}
\label{sec:convergence}

The custom electromagnetic solver is tested on two benchmarks, a dielectric and a weakly absorbing
dispersive medium, using seven grids from 50 to 25 nm and a 22 nm reference grid. The dielectric
benchmark recovers the expected second-order convergence \cite{Taflove2005,JimenezMejia2015}
($p=\ConvergenceOrderDielectric{}$, $R^2=\ConvergenceRSquaredDielectric{}$). The dispersive
benchmark gives a lower effective order of about \ConvergenceOrderDispersive{} and follows a single
power law less closely ($R^2=\ConvergenceRSquaredDispersive{}$). In both benchmarks, the stability
ratio $\rho_i$ of section~\ref{sec:verification} stays within the acceptance limit at every
resolution, so the error remains bounded. The production mesh uses 4 nm cells across the film and
20 nm cells across the waveguide core, both finer than the smallest benchmark grid. These tests
support the numerical stability of the solver. The absolute error of the full three-dimensional
calculation is not quantified. Supplementary S1 gives the complete benchmarks and the mesh
analysis.
\section{Conclusion}
\label{sec:conclusion}

We model the sequential optical switching and readout of an 8 nm GdFeCo film integrated into a
Si$_3$N$_4$ waveguide. At the reference condition, with a film-averaged absorbed fluence of
\ProductionFluence{}, the calibrated model gives a switched fraction of \ToggleSwitchFraction{} from
either saturated state. The switched fraction stays within \AnchorSwitchRange{} over a
\FluenceWindowFactor{}-fold range of incident fluence. Applying the single-cell threshold to the
absorption map reproduces the five three-dimensional results to within $8\times10^{-4}$. Within the
tested range, and in a model whose cells are heated independently, the incomplete switching is set
by the non-uniform absorption of the guided mode. This result provides a quantitative basis for
the explanation proposed for the size-dependent contrast of on-chip Co/Gd switching
\cite{Li2025onchip}. It also suggests that a more uniform guided-mode profile could enlarge the
switched area more efficiently than a higher pulse energy.

When the film starts from zero magnetization, the local moments recover in random directions, and
a fraction \OccupancyMean{} of the cells ends with a negative $x$ component of the FeCo moment. Heating
determines which cells reach the switching regime, and thermal noise selects their orientation when
no initial magnetic direction remains.

The 532 nm probe distinguishes the saturated and stochastic states mainly through its polarization.
Its transmitted, reflected, and absorbed energy fractions vary by at most 0.02 between the states.
A saturated film gives background-subtracted Faraday and Kerr rotations of \FaradaySatOdd{} and
\KerrSatOdd{}, which are about \FaradaySuppression{} and \KerrSuppression{} times smaller,
respectively, in the stochastic state. The reflected field also contains a geometric background of
\KerrPedestal{}, about 1.4 times the saturated magnetic Kerr contribution, and the magnetic signal
can be recovered by subtracting a reference signal. The lobe ratio of the transmitted mode, which
changes from \LobeBalancePlus{} in the saturated state to \LobeBalanceZero{} in the stochastic
state, provides a second predicted signature.

These predictions rest on calibrated switching rules, an empirical permittivity for a homogeneous
film, and phenomenological magneto-optic weights. The model also lacks lateral heat flow and
exchange stiffness between cells, and it does not follow domain coarsening at later times. It
therefore does not predict physical domain sizes or the sharpness of the switched boundary. The
optical readout has yet to be demonstrated experimentally, and the model does not include a
mode-selective detector for the lobe signature. The assumed film permittivity, the magnetic
recovery, and the readout can be tested by film ellipsometry, magnetic imaging, and referenced
waveguide polarimetry, respectively. For such readout experiments, the model identifies the
geometric background of the reflected signal as the main constraint.

\section*{Code availability statement}

The calculations were performed with MagnetoPhotonic.jl v\PackageVersion{}. The versioned source
is archived in Zenodo at
\href{https://doi.org/10.5281/zenodo.21866443}{doi:10.5281/zenodo.21866443}
\cite{MagnetoPhotonicSoftware} and is also available under the
\href{https://github.com/physarief78/MagnetoPhotonic.jl/tree/v0.2.0}{v0.2.0 GitHub tag}. The code is
released under GPL-3.0-only. The script \path{examples/replicate_validation.jl} runs the simulation
and analysis workflow used for the reported figures.

\section*{Data availability statement}

The data supporting this study are available in Zenodo at
\href{https://doi.org/10.5281/zenodo.21874572}{doi:10.5281/zenodo.21874572}
\cite{MagnetoPhotonicData}. The deposit contains the figure-data bundle, analysis and rendering
scripts, pinned Julia environment, provenance information, convergence results, fluence summaries,
and the \SeedCount{} stochastic-realization summaries.

Large raw field files and probe-field movies are omitted because of their size. All arrays used to
generate the figures are included in the deposit, and the package-native field data can be
regenerated from the archived release. The omitted files are available from the corresponding
author on reasonable request. The data and documentation are released under CC BY 4.0, and the Julia
scripts retain GPL-3.0-only.

\section*{Acknowledgements}

This research was carried out at the Department of Physics, Universitas Padjadjaran.

\section*{Declaration of generative AI and AI-assisted technologies in the research and writing
process}

The conceptualization, methodology, analysis, and conclusions of this research are the authors'
own. The authors used Claude (Anthropic; Opus 5 and Opus 5.5) to refactor the research code into
the archived package and to improve the language of the manuscript. The authors reviewed and
verified all AI-assisted output and take full responsibility for the content of this publication.

\section*{Funding}

This research received no specific grant from any funding agency in the public, commercial, or
not-for-profit sectors.

\section*{Conflict of interest}

The authors declare no conflict of interest.

\section*{Author contributions}

\textbf{Muhammad Arief Mulyana}: conceptualization (lead), data curation (lead), formal analysis
(lead), investigation (lead), methodology (lead), software (lead), visualization (lead),
writing -- original draft (lead), and writing -- review and editing (supporting).
\textbf{Budi Adiperdana}: formal analysis (supporting), methodology (supporting), project
administration (equal), supervision (supporting), validation (equal), and writing -- review and
editing (equal). \textbf{Ayi Bahtiar}: project administration (equal), resources (lead),
supervision (lead), validation (equal), visualization (supporting), and writing -- review and
editing (equal).

\bibliographystyle{unsrt}
\bibliography{references}

\end{document}